\documentclass[ ]{copernicus2}

\usepackage{color}
\usepackage[T1]{fontenc}

\begin{document}

\title{On the cosmological vacuum and dark energy
}

\author[1,2]{R. A. Treumann}
\affil[1]{International Space Science Institute, Bern, Switzerland}
\affil[2]{Department of Geophysics and Environmental Sciences, Munich University, Munich, Germany}


\runningtitle{Dark energy}

\runningauthor{R. A. Treumann}

\correspondence{R. A.Treumann\\ (art@geophysik.uni-muenchen.de)}

\received{ }
\revised{ }
\accepted{ }
\published{ }


\firstpage{1}

\maketitle

{\bf{Abstract}}. --  A degenerate fermionic vacuum population is suggested. Based on the abundance of the dark energy density in the Universe the vacuum particle mass and number density are estimated. The obtained mass is in reasonable agreement with observations and predictions of the neutrino mass. Number densities are higher than those of the relativistic relic neutrino component.

\vspace{0.5cm}
\section{Introduction}
Recent astrophysical observations \citep{riess1998,perl1999} tell convincingly that the vacuum is not ``empty", in whatsoever meaning of the word. It contains a substantial energy density that is attributed to some unknown ingredient \citep{zeld1967,zeld1968,turner1998,perl1999a} and its quantum fluctuations. This ``dark energy'' content has negative pressure, which accelerates the universal expansion, and plays the role of an equivalent to a (generalized) cosmological constant  \citep{einstein1917,dirac1937,wilczek1984,weinberg1989,carroll1992}. Negative pressure for a hypothetical real positive mass ($m_v>0$) particle component occupying the vacuum implies that it occupies negative energy states $\epsilon_{v\mathbf{p}}<0$, giving rise to a pressure $P_v=N_v\langle\epsilon_{v\mathbf{p}}\rangle<0$, where $N_v$ is the vacuum density of these hypothetical particles, and the angular brackets $\langle\dots\rangle$ indicate averaging.

Nothing is known about the temperature of the vacuum. In the early universe prior to inflation it might have been very hot but presumably has cooled down to low temperture during inflation. CMB photon and closely related relic neutrion temperatures today are close to absolute zero. Let us assume the vacuum had temperature $T=0$ (in energy units), a temperature far away from any temperature at which any of the high energy quantum field theories would come to work, except of quantum electrodynamics, an assumption that is not in disagreement with any experience or observation. At zero temperature (or close to it, which in cosmological context means at times well after decoupling) the real vacuum in the cosmos is subject to electrodynamics only. It, for instance, implies that we can use electromagnetic waves as carrier of information and can measure everything against the stable electrodynamic vacuum. Another consequence is buried in the constancy of the electrodynamic properties of the vacuum, the constancy of its electric and magnetic susceptibilities $\mu_0$ and $\epsilon_0$ which includes the constancy of the resistance of the vacuum. 

This contemporary electrodynamic vacuum, if containing a hypothetical real particle component (particles of mass, charge, spin), then this component must be subject to Fermi statistics. The only bosons propagating on it are photons and gravitons, the former constituting the electrodynamic fields. Moreover, let it occupy negative energy states, as suggested above and had already been pointed out and proposed by P.A.M. Dirac when discovering his relativistic equation of the electron. However, as it turned out and was interpreted differently in QED, neither electrons nor positrons don't populate the vacuum. 

At negative energy levels $\epsilon_\mathbf{p}=-\sqrt{\mathbf{p}^2c^2+m_v^2c^4}=-\epsilon_{v\mathbf{p}}<0$ and low temperature $T_v\equiv\beta_v^{-1}<m_vc^2$, assumed to be less than the hypothetical particle rest mass energy $m_vc^2$ , the Fermi distribution (in the absence of negative temperatures\footnote{For the non-existence of negative temperatures $T<0$ see \citep{dunkel2014}. In fact, temperatures are defined from kinetic theory through mean square velocity differences $(\overline{\Delta v^2})$ which, for positive occupation numbers of states, are always positive definite.}) implies that the average occupation number of states is $\bar{n}_\mathbf{p}=1$. All negative energy states are occupied. Summed over all states from $-\infty$ to 0 would yield an infinite vacuum particle number, which cannot be true. The number of particles respectively their number density $N_v$ should be finite also in the vacuum. Thus not all negative energy levels can be occupied. Instead the negative fully populated sea is somehow mirror symmetric to the positive domain of particle energies at zero temperature. Inspection of the Fermi distribution for negative energies at $T_v=0$
\begin{equation}
\bar{n}_\mathbf{p}=\Big[\mathrm{e}^{-\beta_v(\epsilon_{v\mathbf{p}}+\mu_v)}+1\Big]^{-1}
\end{equation}
with $\mu_v$ the chemical potential, suggests that the chemical potential in the vacuum must be negative, $\mu_v<0$, if the Fermi distribution should make sense also there. Occupation of negative energy levels is in this case limited by the negative vacuum Fermi energy
\begin{equation}
\mu_v\equiv -\epsilon_{vF}= -\frac{\hbar^2}{2m_v}\Big(\frac{6\pi^2}{g}N_v\Big)^\frac{2}{3}
\end{equation}
where $N_v$ is the unknown and, in principle, not directly measureable vacuum particle density. Levels $\epsilon_\mathbf{p}<\epsilon_{vF}$ are empty, while levels $\epsilon_{vF}\leq\epsilon_\mathbf{p}\lesssim 0$ are occupied with $\bar{n}_\mathbf{p}=1$. 

\section{Estimates}
Little is known about any such vacuum population. The above Fermi energy, a completely unknown quantity, contains two undetermined quantities, the vacuum particle mass, $m_v$, which for real particles is positive, and the vacuum particle number density $N_v$. 

We can, however, impose conditions. The first is that the vacuum Fermi energy, for a sufficiently large number of negative energy states to exist in the vacuum, should at least equal the rest-mass energy
$-\epsilon_{vF}\sim-m_vc^2$ or~equivalently $\epsilon_{vF}\sim m_vc^2$
where $\epsilon_{vF}$ is taken positive. This gives 
\begin{equation}
m_vc^2\lesssim\frac{\hbar^2}{2m_v}\bigg(\frac{6\pi^2}{g}N_v\bigg)^\frac{2}{3}
\end{equation}
Astronomical observations of the accelerated expansion of the universe \citep{riess1998,perl1999} tell that our epoch is characterized by the increasing dominance of the vacuum pressure which drives the cosmological expansion. Our current cosmological epoch is special in the sense that vacuum and baryonic energy densities are of same order of magnitude. The vacuum energy density, however, is constant while the mass density decreases with cosmological expansion. Hence, it increasingly overcomes both the visible as well as dark matter contributions plus the cosmic background radiation pressure including any other radiative components. From these observations a negative vacuum pressure, which in our interpretation corresponds to the negative energy density $-\rho_\Lambda$, has been inferred which seems to be either exactly or at least very close to the value expected for a stationary, i.e. time-independent vacuum. The index $\Lambda$ relates the cosmological vacuum energy density and (the cosmological constant) $\Lambda$ \citep{wilczek1984,weinberg1989,carroll1992}. Writing for the vacuum energy density $N_v\epsilon_{vF}$ and setting it equal to the observed cosmological value we have
\begin{equation}\label{eq-vacen}
N_v\epsilon_{vF}\sim \rho_\Lambda
\end{equation}
Since $\rho_\Lambda=\Omega_\Lambda\rho_{crit}$ is known, with $\rho_\mathit{crit}$ the critical cosmological energy density, the two above equations provide a means of estimating both the number density $N_v$ of particles in the cosmological vacuum and the mass of the hypothetical Fermionic population of the vacuum. 

Define a relative vacuum particle mass $\varpi_v=m_v/m$, where $m$ is the electron mass. Then, from the last equality follows for the vacuum number density 
\begin{equation}
N_v=\bigg(\frac{2m\rho_\Lambda}{\hbar^2}\bigg)^\frac{3}{5}\bigg(\frac{g}{6\pi^2}\bigg)^\frac{2}{5}\varpi_v^\frac{3}{5}
\end{equation}
Inserting this into the condition on the vacuum Fermi energy (\ref{eq-vacen}) yields an estimate for the relative mass of the vacuum Fermions
\begin{equation}
\varpi_v\lesssim \frac{\epsilon_{vF}(\varpi_v)}{mc^2} \sim \frac{\rho_\Lambda}{mc^2N_v}
\end{equation}
which is an implicit equation. Resolving it gives
\begin{equation}\label{eq-muv}
\varpi_v\lesssim\bigg(\frac{3}{4}\frac{\pi^2}{g}\bigg)^\frac{1}{4}\bigg(\frac{\rho_\Lambda\lambda_C^3}{mc^2}\bigg)^\frac{1}{4}\approx 1.34\bigg(\frac{\rho_\mathit{crit}\lambda_C^3}{mc^2}\bigg)^\frac{1}{4}
\end{equation}
where the Compton wavelength of an electron $\lambda_C=\hbar/mc$ has been introduced. The central part of this expression depends only on the one free parameter $\Omega_\Lambda$ which is subject to determination making use of astronomical data. $\Omega_\Lambda\approx 0.7$ and $g= 1.75$ for Fermions gives the numerical factor. This is independent on the number density $N_v$. Hence, it can be used to obtain a limit on $\varpi_v$. Inserting into the above expression for the vacuum density it also permits obtaining an important independent vacuum density estimate  
\begin{equation}
N_v\lesssim1.1\bigg(\frac{g}{6\pi^2}\bigg)^\frac{1}{4}\bigg(\frac{\rho_\Lambda}{mc^2\lambda_C}\bigg)^\frac{3}{4}\approx 0.85\bigg(\frac{\rho_\mathit{crit}}{mc^2\lambda_C}\bigg)^\frac{3}{4}
\end{equation}
which may be considered as an upper bound. The critical mass density in the Universe is $\rho_m=9.21\times10^{-27}$ kg/m$^3$ which, in terms of the electron rest mass energy $mc^2$, yields a critical energy density of $\rho_\mathit{crit}/mc^2\approx 1.01\times10^4$. This number gives a rest mass
\begin{equation}
m_v=\varpi_vm_e\sim 5.05\times 10^{-9}m_e
\end{equation}
of the hypothetical particles that make up the vacuum population. Using the rest mass energy of an electron, this becomes
\begin{equation}
m_v\lesssim 2.6 \times 10^{-3}~~\mathrm{eV}/c^2
\end{equation}
a small number which  indicates that the massive vacuum Fermions should be light. With this small mass the vacuum number density is found to be of the order 
\begin{equation}
N_v\lesssim 1.75\times10^{12}~~\mathrm{m}^{-3}
\end{equation}
Such a fermionic vacuum number density is surprisingly high when compared with the densities of interstellar, galactic or even intergalactic space. It is also four orders of magnitude larger than the relativistic relic neutrino background density $N_{\nu}^\mathit{rel}\approx 3.3\times 10^8$ m$^{-3}$ in the Universe. Such a vacuum is quite dense though being very dilute when compared with a vacuum composed of Planckian particles of mass $m_\mathrm{Pl}\sim 10^{19}$ GeV/$c^2$.  

Lower limits on both these numbers are currently not know theoretically. Considering the experimental inference on $\Omega_\Lambda$ we note that the ever more precise measurements have decreased its value slowly from $\Omega_\Lambda\gtrsim 0.7$ to its presently best value $\Omega\approx 0.683$. In order to be on the very conservative side we may thus boldly assume that, at least to current intelligence, an absolute lower observational limit will not be less than $\Omega_\Lambda=0.6$. This gives a factor of $1.29 $ instead of $1.34$ in (\ref{eq-muv}), restricting the vacuum particle mass to the narrow range 
\begin{equation}\label{eq-munuest}
2.5\times 10^{-3} < m_v \lesssim 2.6\times 10^{-3}~~\mathrm{eV}/c^2
\end{equation}
The vacuum number density of the hypothetical particles is, by the same reasoning, subject to the bounds
\begin{equation}
1.56\times10^{12}< N_v\lesssim 1.75\times10^{12}~~\mathrm{m}^{-3}
\end{equation}
These densities are the famous disturbing $\sim110$ orders of magnitudes below a hypothetical vacuum number density based on the assumption of a vacuum built of planckions. In terms of mass ratios between planckions of mass $m_\mathrm{Pl}$ and the vacuum particles this reduces to a ``somewhat fairer''  $m_\mathrm{Pl}/m_v\sim30$ orders of magnitude discrepancy. If the current theory has anything in common with reality then planckions are ruled out by it at least for the contemporary vacuum as it is experienced over all the astronomical observational times in our past-inflation epoch.

\section{The question for the particles}
Not many Fermions this light are known to us. The most promising particles are neutrinos \citep{hewett2012} of which we know that they interact only weakly with matter. The combined current limits from cosmological probes \citep{bellazzini2016,cuesta2016} so far available are upper bounds $m_\nu=\sum_i m_{\nu_i}< 0.12$ eV/c$^2$ which do not rule out that the real vacuum neutrino mass would be somewhat lower. Neutrino oscillations of real (i.e. non-vacuum) relativistic non-degenerate neutrinos indicate non-vanishing mass differences $|\Delta m_\nu|^2$ between the three different generations of neutrinos. In contrast to the free relativistic neutrino background and neutrinos produced in the Universe in $\beta$ decay and other elementary interactions, vacuum neutrinos are degenerate at $T_v\approx 0$. Theory suggests that the masses of the three degenerate neutrino generations are of same order of magnitude $m_\nu\sim |\Delta m_\nu|\sim 5\times10^{-2}$ eV/$c^2$. Observations (on the $2\sigma$ confidence level) indicate $\Delta m_\nu^2\gtrsim 8\times 10^{-5}$ (eV/c$^2$)$^2$ yielding an upper limit of $m_\nu\sim 10^{-2}$ eV/$c^2$  \citep{hewett2012} somewhat larger than our estimated vacuum particle mass range. When measurements will be pushed down to $< 0.1$ eV/$c^2$, for instance by KATRIN \citep{mertens2016}, this will soon be in the reach to become checked.

The theoretical normal hierarchical scheme currently predicts  $m_\nu\lesssim5\times 10^{-3} $ eV/$c^2$ for the electron neutrino. This is close to our completely independent estimate (\ref{eq-munuest}). 

The usual notion that the vacuum is not empty but hosts quantum fluctuations, i.e. virtual particle-antiparticle pairs, would suggest that the degenerate vacuum consist of pairs of virtual neutrinos which are their own antiparticles. Such neutrinos satisfy Majorana statistics instead of Dirac statistics and are known as Majorana neutrinos. Current degenerate experimental upper bounds for Majorana neutrino masses are $m_\nu\lesssim 0.1$ eV/$c^2$ \citep{hewett2012}, which is an upper limit not in disagreement with our estimate (\ref{eq-munuest}). 

Given the independent assumptions made in our estimates, one may become inclined to identify the hypothetical vacuum particle mass with the lightest (electron) neutrino mass 
\begin{equation}
m_v\approx m_{\nu_e}
\end{equation}
of degenerate positive mass neutrinos populating the vacuum and being in negative energy states thus contributing a negative vacuum pressure. For a continuum of vacuum energy levels these neutrinos form a degenerate Landau-Fermi fluid \citep{lif1980}. Such a fluid is subject to quantum fluctuations, which at the low vacuum temperatures propagate as low intensity zero sound at sound speed $c\geq u_{0}\gtrsim p_{vF}/m_v$. Since $p_{vF}\sim m_vc$, one concludes that 
\begin{equation}
u_{0}= c 
\end{equation}
Wavenumbers $k_0<m_vc/\hbar$ imply wavelengths $\lambda_0\gtrsim 18$ Mpc, which is larger than $>$\,0.1\% of the radius of the  Universe and exceeds the diameter of all galaxy clusters. On shorter scales this vacuum can be considered as completely homogeneous. Thus vacuum fluctuations would be interpreted as a spectrum of large-scale zero sound fluctuations propagating at the velocity of light. Longest ``visible wavelengths" would be of the order of the diameter of the visible Universe.  

The vacuum energy density has not changed over the entire time of existence of the vacuum as we know it after the end of inflation. For the same reason the zero-sound speed has been constant over this period. Its constancy implies the constancy of the velocity of light as a property of the invariable vacuum. 

Inverting the argument one may conjecture that the velocity of light is a constant of nature because it is a constant property of the degenerate vacuum which has been invariant for the accessible lifetime of the Universe. 

{The question on the nature of the particles is related to the question whether they are real or virtual. It is usually claimed that such a population consists of virtual particles only. Virtual particles have lifetimes $\tau_v\sim2\pi\hbar/m_vc^2$. With our estimate 
\begin{equation}
\tau_{v}<10^{-12} ~~\mathrm{s}
\end{equation}
is very short. It would require a continuous production of the light vacuum particle component with production rate constant over the complete lifetime of the universe as suggested by the constancy of the dark energy. Whether such a mechanism exists is not known in particularly not at the assumed low contemporary vacuum temperatures. The vacuum particle component should thus rather be real as than virtual, as assumed here, having been generated sometimes in the early universe and remained grossly unchanged since. The quantum fluctuations such a degenerate negative energy vacuum fluid exhibits are not to be traced back to virtual particles. They, as in Landau theory, are zero-point fluctuations in the degenerate quantum fluid of the real vacuum neutrino population.}
 
\section{Cosmological consequences}
The visible Universe  contains a non-degenerate relativistic dilute neutrino population of cosmological very-early universal origin and current temperature $T_\nu/k_B\approx 1.9$ K. These neutrinos have decoupled at roughly $t\sim 1$ s after the Big Bang when they were of high density the order of $10^{25}$ m$^{-3}$ or more and constituted the bulk of neutrinos, becoming gradually diluted in the universal expansion. {In spite of their current density of $1.13\times 10^8$ m$^{-3}$ per species, they do (probably) not contribute substantially to the unknown composition of dark matter. Their todays density is only a small fraction of the density of their hypothetical vacuum sister neutrino component. Moreover, being of positive energy they are unrelated to dark energy. As real positive energy particles they don't generate any negative vacuum pressure.}

The origin of the hypothetical vacuum neutrino component is unknown. During inflation at times $t\lesssim 10^{-32}$ s, they just formed the lowest energy branch of the neutrino energy distribution. One may speculate that at that time they become trapped by the vacuum by some unknown mechanism. Afterwards, over the accessible lifetime of the universe, their density did not change anymore, at least not appreciably, because the inflationary vacuum expansion was terminated. As vacuum population, they escaped our accessibility while providing the negative vacuum pressure. What in the course of inflation enabled the vacuum to trap the {tiny minority of cold} neutrinos {of the total neutrino population}, is unknown.

One might wonder whether today the vacuum neutrinos, in addition to providing the dark energy density, could contribute to the unknown composition of Dark Matter. This is not the case, however, as can be shown by asking up to what radial distance $R$\, from a large local mass $M= \alpha M_\odot$, with $M_\odot$ the solar mass, the local gravitational energy density and thus gravitational attraction could overcome the dark energy pressure of the light degenerate vacuum neutrinos. This leads to
\begin{equation}
R< \frac{GM_\odot mN_v}{\Omega_\Lambda\rho_mc^2}\, \alpha\, \varpi_v \approx 10^{-9}\alpha~\mathrm{m}
\end{equation}
Even the largest clusters of galaxies hosting $\alpha\sim 10^{14}$ solar masses, if concentrated within one point, would dominate the dark energy merely within a sphere of $\sim$ 100 km radius from their centers. A light neutrino vacuum can thus barely be made responsible for the Dark Matter content of the Universe.


\begin{thebibliography}{99}
\bibitem[Bellazzini et al.(2016)]{bellazzini2016} {Bellazzini B, Csaki C, Hubisz J,  Serra J, and Terning J.}:~{J High Energy Phys.} {6},{104, doi: 10.1007/JHEP06(2016)104}, {2016}.

\bibitem[Carroll et a.(1992)]{carroll1992} {Carroll S.M., Press W.H. and Turner E.L.}: {Ann. Rev. Astron. Astrophys.} {30}, {499, doi: 10.1146/annurev.aa.30.090192.002435}, {1992}.

\bibitem[Cuesta et al.(2016)]{cuesta2016} {Cuesta A.J., Niro V. and Verde L.}: {Phys. Dark Universe} {13}, {77, doi: 10.1016/j.dark.2016.04.005},  {2016}.

\bibitem[Dirac(1937)]{dirac1937} {Dirac P.A.M.}: {Nature} {139}, {323, doi: 10.1038/139323a0,} {1937}.

\bibitem[Dunkel and Hilbert(2014)]{dunkel2014} {Dunkel J. and Hilbert S.}: {Nature Phys.} {10}, {67, doi: 10.1038/nphys2815}, {2014}.

\bibitem[Einstein(1917)]{einstein1917} {Einstein A.}: {Sitzungsber. K\"onigl. Preuss. Akad. Wiss., Phys.-Math. Klasse}, {142}, {8. Februar 1917}.

\bibitem[Hewett and Weerts(2012)]{hewett2012} {Hewett J. L. and Weerts H. (eds.)  et al.}:~{Fundamental Physics at the Intensity Frontiers, Report of Workshop held December 2011 in Rockville, MD, Chpt. 4: Neutrinos}~(ANL-HEP-TR-12-25, SLAC-R-991, Rockville, MD 2012).

\bibitem[Landau et al.(1980)]{lif1980} {Landau~L.D.,~Lifschitz~E.M.~and~Pitaevskii L.P.}: Statistical Physics, Part 2 (Butterworth-Heinemann, Oxford UK, 1998).

\bibitem[Mertens(2016)]{mertens2016} {Mertens S.}: {J. Phys.: Conf. Series} {718}, {022013, doi: 10.1088/1742-6596/718/2/022013},  {2016}.


\bibitem[Perlmutter et al.(1999a)]{perl1999} {Perlmutter S. et al.}: {Astrophys. J.} {517}, {565, doi: 10.1086/307221}, {1999}.

\bibitem[Perlmutter et al.(1999b)]{perl1999a} {Perlmutter S., Turner M. S. and White M.}: {Phys. Rev. Lett.} {83}, {670, doi: 10.1103/PhysRevLett.83.670}, {1999}.

\bibitem[Riess et al.(1998)]{riess1998} Riess A.G. et al.: {Astron. J.} {116}, {1009, doi: 10.1086/300499}, {1998}.

\bibitem[Turner(1998)]{turner1998} {Turner M.S.}: {Dark matter and dark energy in the universe}, arXiv:astro-ph/9811454, 1998.


\bibitem[Weinberg(1989)]{weinberg1989} {Weinberg S.}: Rev. Mod. Phys. 61, 1, doi: 10.1103/RevModPhys.61.1, 1989.

\bibitem[Wilczek(1984)]{wilczek1984} {Wilczek F.}: {Phys. Reports} {104}, {143, doi: 10.1063/1.2780139,} {1984}.

\bibitem[Zeldovich(1967)]{zeld1967} {Zeldovich Y.B.}: {JETP Lett.} {6}, {316}, {1967}.

\bibitem[Zeldovich(1968)]{zeld1968} {Zeldovich Y.B.}: {Sov. Phys. Usp.} {11}, {381}, {1968}.

\end{thebibliography}
\end{document}